\documentclass[reprint,nofootinbib,amsmath,amssymb,aps]{revtex4-2}

\usepackage{float}
\usepackage{xcolor}
\usepackage{graphicx}
\usepackage{hyperref}
\usepackage{dsfont}
\usepackage[normalem]{ulem}

\graphicspath{{img/}}

\renewcommand{\Re}{\mathrm{Re}}

\newcommand{\SixJ}[2]{\left\lbrace \begin{array}{ccc} #1 \\ #2 \end{array}\right\rbrace}

\begin{document}

\title{Tunneling of quantum geometries in spinfoams}

\author{Pietro Don\`a}
\email{pietro.dona@cpt.univ-mrs.fr}
\affiliation{Aix-Marseille Univ, Universit\'e de Toulon, CNRS, CPT, Marseille, France}
\author{Hal M. Haggard}
\email{hhaggard@bard.edu}
\affiliation{Physics Program, Bard College, 30 Campus Road, Annandale-on-Hudson, NY 12504, USA}
\author{Carlo Rovelli}
\email{crovelli@uwo.ca}
\affiliation{Center for Space, Time and the Quantum, 132UNCS88 Marseille, France}
\affiliation{Dept.\,of Philosophy and Rotman Institute, Western University, N6A\,3K7, London ON, Canada}
\affiliation{Aix-Marseille Univ, Universit\'e de Toulon, CNRS, CPT, Marseille, France}
\affiliation{Perimeter Institute, 31 Caroline Street North, N2L\,2Y5 Waterloo ON,  Canada}

\author{Francesca Vidotto}
\email{fvidotto@uwo.ca}
\affiliation{Dept.\,of Physics \& Astronomy, Western University, N6A\,3K7, London ON, Canada}
\affiliation{Dept.\,of Philosophy and Rotman Institute, Western University, N6A\,3K7, London ON, Canada}

\date{\today}

\begin{abstract}
\noindent Quantum gravitational tunneling effects are expected to give rise to a number of interesting observable phenomena, including, in particular, the evolution of black holes at the end of their existence or the emergence of the early universe from a quantum phase.
Covariant Loop Quantum Gravity provides a framework to study these phenomena, yet a precise identification of tunneling processes is still not known. Motivated by this question, we consider a related, simpler case, that of Ponzano-Regge amplitudes: we find a detailed analogy of a class of simple transition amplitudes with tunneling processes in non-relativistic quantum mechanics.
\end{abstract}

\maketitle    


\section{Introduction}
\label{sec:intro}
Predicted corrections to general relativity of first order in $\hbar$ are generally too small to be observed. An alternative strategy to find signatures of quantum gravity is to search for processes forbidden in the classical theory, such as quantum gravitational tunneling. A concrete realization of spacetime tunneling, attracting attention recently because of its phenomenological implications, is the quantum transition of a black hole to a white hole, which could happen at the end of the Hawking evaporation \cite{Rovelli:2014cta, Rovelli:2017zoa, Haggard:2014rza, DeLorenzo:2015gtx, Rovelli:2018cbg, Rovelli:2018okm, Bianchi:2018mml, Ashtekar2018, Lewandowski:2022zce, Giesel:2022rxi, Kazemian:2022ihc, Han:2023wxg}. This proposal joins a longstanding interest in the quantum bounce, which, in Loop Quantum Gravity, is expected to replace the classical Big Bang singularity
\cite{AshtekarEtAl2006, Ashtekar2006quantum, Bianchi:2010zs,Vidotto:2011qa,Han:2024ydv}, potentially with a tunneling phenomenon \cite{Asante:2021phx,Dittrich:2023rcr,Bojowald:2021cqg,Motaharfar:2022pjp}.

Covariant Loop Quantum Gravity is a background-independent, Lorentzian, sum-over-histories quantization of general relativity perfect for studying these phenomena. In particular, the recent progress in the computation of the black-to-white hole tunneling \cite{Christodoulou:2016vny, DAmbrosio:2020mut, Christodoulou:2018ryl, Soltani:2021zmv, Christodoulou:2023psv} have highlighted some conceptual and technical difficulties, among them: defining what tunneling means in a theory of spacetime, specifying suitable boundary data, computing the Lorentzian spinfoam amplitude, finding meaningful physical quantities to calculate, and interpreting the results. Spinfoams are the spacetime analogs of Feynman graphs, and frequently covariant Loop Quantum Gravity is referred to as spinfoam theory.

This work focuses on the meaning of quantum geometric tunneling and how this manifests in a spinfoam theory. We study a specific transition amplitude in the relatively simple context of the Ponzano-Regge spinfoam model and how we can interpret this process as the tunneling of quantum geometry. The Ponzano-Regge model is the most straightforward example of a spinfoam theory. It describes Euclidean quantum gravity in three dimensions. This choice isolates some of the open conceptual questions and avoids the complications of the full theory. Our synthesis of results in the literature allows us to detail several analogies with the tunneling of a point particle through a potential barrier in the path integral formulation of quantum mechanics.

\section{The Ponzano-Regge spinfoam theory}
The Ponzano-Regge spinfoam theory is a path integral quantization of three-dimensional (3D) Euclidean gravity. It is regularized on a simplicial 2-complex and assigns transition amplitudes to three-valent spin network states at the boundary. The states of the theory describe two-dimensional quantum surfaces discretized with Euclidean triangles dual to the nodes of the boundary spin networks. The spinfoam amplitude provides the quantum dynamics of these surfaces and decomposes into local amplitudes associated with the vertices of the 2-complex, which are dual to tetrahedra. We assume that the reader is familiar with the basic concepts of this theory. We summarize the few elements we need here, and refer to the original work \cite{Regge:1968} and the literature for more in-depth presentations (e.g., see \cite{Roberts:1998zka,Freidel:2004vi, Freidel:2005bb,Barrett:2008wh,Rovelli:2014ssa} and references therein).

\begin{figure}
    \centering
    \includegraphics[width=0.7\linewidth]{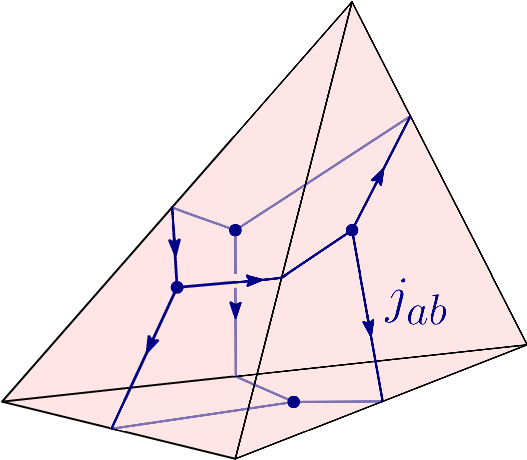}
    \caption{The tetrahedral spin network at the boundary of a Ponzano-Regge vertex. Each node is dual to a boundary triangle of the tetrahedron.}
    \label{fig:spinnetwork}
\end{figure}

At the boundary of a spinfoam vertex, we find the tetrahedral spin network depicted in Fig.~\ref{fig:spinnetwork}. We label the four nodes (dual to each triangle) with $a=1,\cdots,4$ and the oriented links with an oriented couple $ab$ and a spin $j_{ab}$. 
The spins are eigenvalues of the length squared operator $J_{ab}^2$ and give the length of the tetrahedral edge the spin network link crosses
\begin{equation}
\label{eq:spectrum}
\ell_{ab} = \hbar G \sqrt{j_{ab}(j_{ab}+1)} \ ,
\end{equation}
with $j_{ab}$ a half-integer.
For the remainder, we work in units with the 3D Planck length set to one, ($\ell_P=\hbar G=1$).

The Ponzano-Regge vertex amplitude in the spin network basis is given by the Wigner $\{6j\}$ symbol \cite{Varshalovich:1988krb}
\begin{equation}
    \label{eq:Avspins}
    A_v (j_{ab})=\SixJ{j_{12}&j_{13}&j_{14}}{j_{34}&j_{24}&j_{23}} \ .
\end{equation}
A general transition amplitude provides the dynamics of a quantum surface. These amplitudes are given by products of the local amplitudes, $A_v$, summed over the intermediate quantum numbers, which implements the path integral sum over histories.

\section{The classical evolution}
The path integral of ordinary quantum mechanics is dominated, in the semiclassical limit, by classical paths. Classical paths are solutions of the equations of motion of the  underlying classical theory, and the transition amplitude between an initial and final state ($x_0$ and $x_1$) reduces to
\begin{equation}
\label{eq:pathintegral}
    \int_{x_0}^{x_1} \mathcal{D}[x] e^{i S[x]} \approx \sqrt{\frac{i}{2\pi}\frac{\partial^2 S[x_c]}{\partial x_0\partial x_1}} e^{iS[x_c]} \ ,
\end{equation}
where $S[x_c]$ is Hamilton's principal function, i.e. the action evaluated on a solution of the equations of motion, $x_c(t)$, compatible with the boundary conditions.

A similar scenario is realized in the Ponzano-Regge spinfoam theory. The underlying classical theory is three-dimensional Regge calculus, a discrete version of General Relativity in Euclidean spacetime. This section briefly reviews the key features of Regge calculus we need in this work.

In the first-order formulation \cite{Barrett:1994nn, Bahr:2009qd}, the fundamental variables are the lengths of the edges of the triangulation and the dihedral angles between two triangles sharing that length in a tetrahedron. The edge lengths and their dual angles are symplectically conjugate. The equations of motion fix the angles as functions of the lengths to be the dihedral angles of an Euclidean tetrahedron (see Appendix~\ref{app:geo}). They also require that the dihedral angles around a bulk length sum to $2\pi$, ensuring the flatness of 3D gravity. 

Viewed as a canonical theory, Regge calculus describes the evolution of two-dimensional surfaces using Hamilton's principle function \cite{Rovelli2004, Rovelli:2014ssa, Dittrich:2011ke}. Each surface comprises a collection of triangles glued together along their edges. The lengths of the edges of all triangles must satisfy triangle inequalities to ensure they exist. This is a constraint on the boundary data of the theory. At the quantum level, three-valent spin networks represent quantum surfaces, and triangle inequalities are required by $SU(2)$ gauge invariance at the nodes (dual to triangles). The dynamics of classical three-dimensional canonical gravity can be encapsulated as a series of local moves gluing \textit{Euclidean tetrahedra} to a surface (in all possible ways) \cite{Dittrich:2011ke}. The corresponding Hamilton function,
\begin{equation}
    \label{eq:HamiltonRegge}
    S_R(\ell_{ab})= \sum_{ab} \psi_{\!ab}\!(\ell_{ab})\, \ell_{ab} \ ,
\end{equation}
is the Regge action evaluated on a flat Euclidean tetrahedron with lengths $\ell_{ab}$ and external dihedral angles $\psi_{ab}$, and is the fundamental building block of all discrete classical solutions.

\begin{figure}[htbp] 
    \raisebox{-0.5\height}{\includegraphics[width=0.45\linewidth]{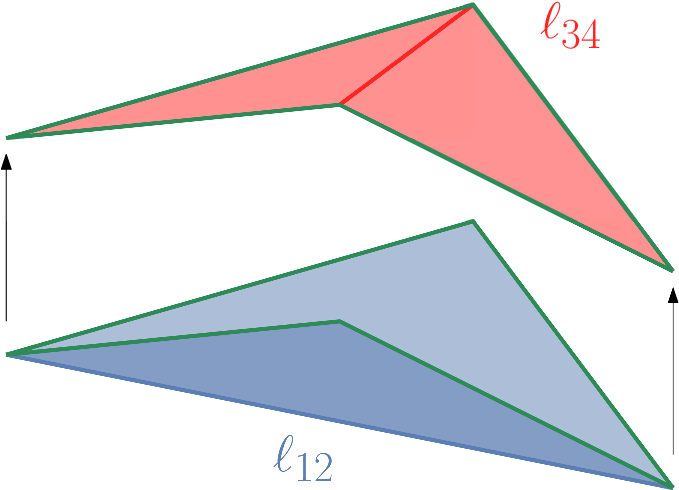}}
    \hspace{0.2in}
    \raisebox{-0.5\height}{\includegraphics[width=0.45\linewidth]{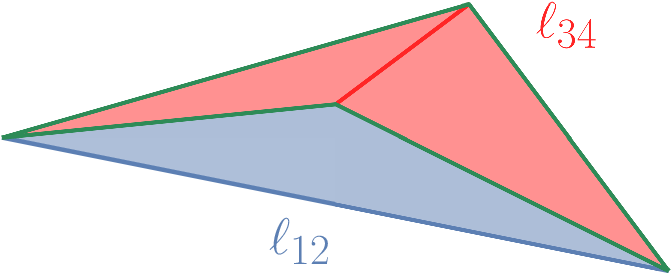}}
\caption{\emph{Left panel.} A simple canonical evolution of a surface discretized with two triangles (bottom pair, in blue) to a surface discretized with two triangles (top pair, in red). \emph{Right panel.} We evolve the bottom surface into the top surface by gluing them to form an Euclidean tetrahedron.
\label{fig:transition}}
\end{figure}

For concreteness, we focus on one specific example: the evolution of a surface discretized by two triangles sharing one edge into a surface discretized by two triangles sharing a different edge (left panel of Fig.~\ref{fig:transition}). The two surfaces share four boundary lengths that we assume are fixed once and for all. The remaining two lengths $\ell_{12}$ and $\ell_{34}$ completely characterize the two surfaces. We evolve one surface into the other by gluing in an Euclidean tetrahedron (see right panel of Fig.~\ref{fig:transition}).

Not all surfaces characterized by triangle-allowed values of $\ell_{12}$ and $\ell_{34}$ are compatible with classical canonical evolution. 
Fortunately, an elegant and efficient criterion uses the tetrahedral volume to distinguish between classically allowed and classically forbidden evolution. 

The squared volume of a Euclidean tetrahedron, $V^2$, can be computed directly from the lengths of its edges and their connectivity using the Cayley–Menger determinant. The formula is detailed in Appendix~\ref{app:geo}. An ordered set of edge lengths $\{\ell^e\}$ form a Euclidean tetrahedron if and only if they satisfy the triangle inequalities for each of the faces and satisfy $V^2(\ell^e)>0$. The corresponding evolution is classically forbidden if $V^2(\ell^e)<0$.

\begin{figure}
    \centering
    \includegraphics[width=0.8\linewidth]{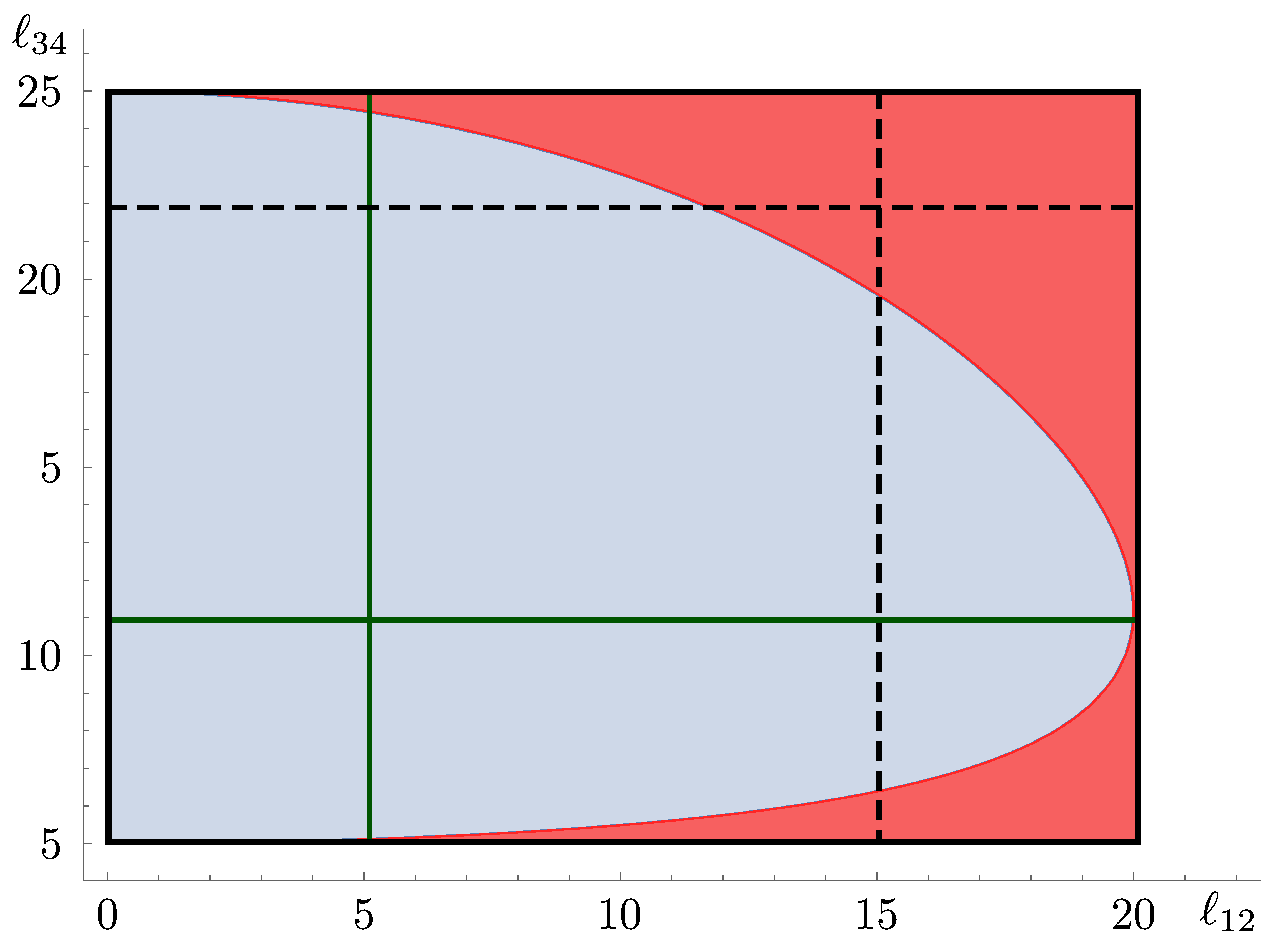}
    \caption{The configuration space of the initial and final surfaces in terms of $\ell_{12}$ and $\ell_{34}$  for fixed lengths $\ell_{13} = \ell_{23} = 10$ and $\ell_{14} = \ell_{24} = 15$. We color in blue the classically allowed region ($V^2(\ell_{ab})>0$), and in red the classically forbidden region ($V^2(\ell_{ab})<0$). The green solid lines and dashed black lines are examples of initial and final surfaces compatible with classically allowed and classically forbidden evolution, respectively.}
    \label{fig:region1}
\end{figure}

In Fig.\ref{fig:region1}, we visualize the evolution of surfaces in the $\ell_{12}$ and $\ell_{34}$ configuration space, while keeping other lengths fixed ($\ell_{13} = \ell_{23} = 10$ and $\ell_{14} = \ell_{24} = 15$). Each value of $\ell_{12}$ represents an initial surface, and each value of $\ell_{34}$ represents a final surface. All surfaces satisfying triangle inequalities fit within a rectangle (pictured here by the darkened border).

Vertical lines in Fig.\ref{fig:region1} represent any possible evolution of the initial surface, while horizontal lines represent any possible (backward) evolution of the final surface. In blue, we highlight the classically allowed evolution region associated with an Euclidean tetrahedron with $V^2 > 0$. Conversely, the classically forbidden region, associated with a tetrahedron with $V^2 < 0$, is depicted in red.

Specific pairs of hypersurfaces, such as those with $\ell_{12}=5$ and $\ell_{34}=11$ (the solid lines of Fig.~\ref{fig:region1}), intersect within the classically allowed region. This intersection implies the possibility of classical evolution, enabling one surface to transition into the other via an Euclidean tetrahedron.

In contrast, pairs of surfaces like those with $\ell_{12}=15$ and $\ell_{34}=22$ (dashed in Fig.~\ref{fig:region1}) intersect in the classically forbidden region, indicating the absence of classical evolution connecting the latter pair of surfaces.

Does the intersection of surfaces within the classically forbidden region correspond to another viable but non-classical geometry? Indeed, it does. Observe that if we use the same expression for dihedral angles (momenta) as a function of the lengths, Eq.~\eqref{eq:angles}, it is necessary to analytically continue these angles to purely imaginary values (as $V^2(\ell_{ab})<0$ in \eqref{eq:angles}), see \cite{Barrett1994semiclassical, Roberts:1998zka}.

Also, note the impossibility of embedding the two surfaces in Euclidean space since $V^2(\ell_{ab})<0$. Nonetheless, by complexifying the dihedral angles, we can successfully embed the surfaces as a tetrahedral region of Minkowski space, with signature $(-,+,+)$. This complexification arises from the need to take dot products across the light cone, as discussed in \cite{Sorkin2019, Asante2021}.

\section{The quantum transition amplitude}
In the quantum theory, transition amplitudes characterize the evolution of quantum states. In the example we are studying, the boundary state is the tetrahedral three-valent spin network, depicted in Fig.~\ref{fig:spinnetwork}. The relationship between spins $j_{ab}$ and lengths is defined by \eqref{eq:spectrum}. At the lowest order, the transition amplitude is determined by a Ponzano-Regge vertex amplitude, as specified in \eqref{eq:Avspins}. In the semiclassical regime, characterized by large quantum numbers, the amplitude \eqref{eq:Avspins} has been extensively studied \cite{Regge:1968,Roberts:1998zka,Dupuis:2009sz,AquilantiEtAl2012,Dona:2019jab}.

If the evolution between the boundary surfaces is classically allowed (the surfaces form a Euclidean tetrahedron with $V^2(\ell_{ab}>0)$), the Ponzano-Regge transition amplitude for large quantum numbers is well approximated by 
 \begin{equation}
 \label{eq:semicalssical}
 A_v \approx \tfrac{1}{2\sqrt{12\pi V(\ell_{ab})}} \exp\left(i {S}_R[\ell_{ab}] + i\frac{\pi}{4}\right) + c.c. \ ,
\end{equation}
where we interpret the Regge action of the Euclidean tetrahedron ${S}_R[\ell_{ab}]$ as Hamilton's principal function computed on a solution of the equation of motion of Euclidean Regge calculus in three dimensions. This is the spinfoam equivalent of \eqref{eq:pathintegral}, where the path integral is dominated by the classical paths compatible with the boundary data.

What if the classical evolution of the boundary data is forbidden? The spinfoam transition amplitude is still not vanishing. In the semiclassical regime, we can approximate the Ponzano-Regge vertex amplitude with the asymptotic expression\cite{Regge:1968,AquilantiEtAl2012}\footnote{This formula was discussed in the original paper, and we report it as is. We plan to re-derive it in the future using coherent states-based asymptotic analysis techniques \cite{Dona:2017dvf, Dona:2020yao}.}
 \begin{equation}
 \label{eq:PRlorentzian}
 A_v \approx \frac{1}{2\sqrt{12\pi |V(\ell_{ab})|}} \exp\left(-S_R^c[\ell_{ab}]\right)
\end{equation}
where $S_R^c[\ell_{ab}] = -i S_R[\ell_{ab}]$ is the analytic continuation of the Regge calculus Hamilton function for a classically forbidden evolution. The (Euclidean) dihedral angles are analytically continued to complex values with positive imaginary parts. Therefore, $\Re\, S_R^c[\ell{ab}]>0$, and the resulting amplitude $A_v$ is exponentially suppressed, as expected for a classically forbidden process. The volume is purely imaginary, and the extra imaginary unit cancels the $i \pi/4$ phase of \eqref{eq:semicalssical}. 

This is the perfect example of a tunneling process of quantum geometries in spinfoam theories. Evolution between two surfaces (states), which is classically forbidden, can be realized quantum mechanically. The process is rare, as the transition amplitude is exponentially suppressed. The path integral is dominated by the analytic continuation of a solution of the classical equations of motion, and the analytic continuation to the classically forbidden trajectory of Hamilton's principal function characterizes the suppression. The next section discusses the various analogies with more familiar tunneling scenarios in quantum mechanics.

\begin{figure}
    \centering
    \includegraphics[width=0.8\linewidth]{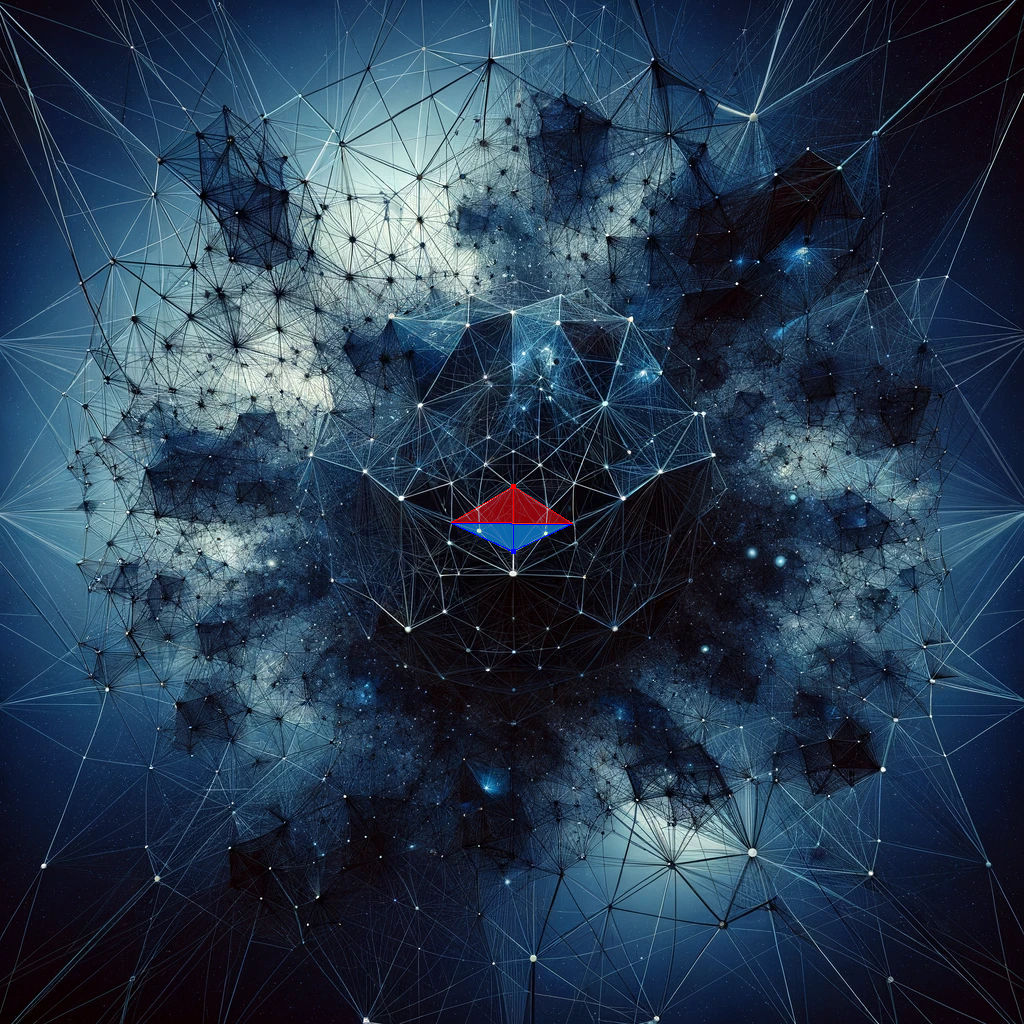}
    \caption{Pictorial representation of a local tunneling bubble. We highlight the initial and final surfaces of the bubble in blue and red. Illustration generated with ChatGPT 4.0.}
    \label{fig:TunnelingBubble?}
\end{figure}

As a concluding remark, we note that both surfaces involved in a tunneling event are legitimate Euclidean surface triangulations. The evolution after or before a tunneling event does not necessarily exhibit anything unusual and can adhere to entirely classical dynamics. In this sense, the tunneling event can be viewed as a completely isolated bubble within an otherwise standard Euclidean evolution, as artistically illustrated in Fig.~\ref{fig:TunnelingBubble?}.

\section{Discussion}
In conclusion, we analyze the analogies, effectively serving as a dictionary, between the tunneling of quantum geometries and the well-understood phenomenon of a point particle tunneling through a potential barrier in quantum mechanics.

\begin{figure}
    \centering
    \includegraphics[width=0.6\linewidth]{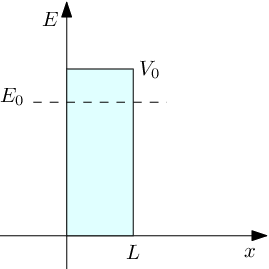}
    \caption{A one-dimensional potential barrier.}
    \label{fig:barrier}
\end{figure}

Consider a point particle of mass $m$ in one dimension impinging on a potential barrier of width $L$ and height $V_0$ (see Fig.~\ref{fig:barrier}). We will assume the particle has fixed energy $E_0$.  

Can the particle traverse the barrier? The kinetic energy of the particle in the region inside the barrier is 
\begin{equation}
    \label{eq:cltrajectory}
    E_{kin} = \frac{p^2}{2m} = E_0 - V_0 \ .
\end{equation}
We encounter two main scenarios. In the first scenario, where the particle's energy exceeds the barrier height ($E_0 > V_0$), the particle possesses sufficient energy to go over the barrier without unusual phenomena. 
The transition amplitude in the semiclassical limit $\hbar\ll 1$ is dominated by the classical trajectory over the barrier.
The transition amplitude is dominated by the classical action evaluated on the classical equation of motion
\begin{equation}
A({E_{kin}>0})\propto e^{\frac{i}{\hbar} L\sqrt{\frac{m}{2}(E_{0}-V_{0})}} \ .
\end{equation}

In contrast, when the particle's energy is less than the barrier height ($E_0 < V_0$), it lacks the requisite energy to cross the barrier conventionally. However, quantum mechanics allows a tiny probability for the particle to tunnel through the potential barrier. 
The momentum of the particle within the barrier is ill-defined. However, we can define it by analytic continuation to purely imaginary values $p = i\sqrt{2m|(E_0 - V_0)|}$. The transition is classically forbidden. There are no classical paths that dominate the path integral in the semiclassical limit, and the amplitude is exponentially suppressed 
\begin{equation}
A({E_{kin}<0})\propto e^{- \frac{1}{\hbar}L\sqrt{\frac{m}{2}|E_{0}-V_{0}|}}\ .
\end{equation}
The suppression is regulated by the classical action evaluated on the analytic continuation of a classical solution (with imaginary momentum). 

\smallskip 

The parallelism between a point particle penetrating a potential barrier and the tunneling of Euclidean three-dimensional geometries is striking. The volume squared of the tetrahedron plays the role of the particle's kinetic energy. Depending on the volume squared or the kinetic energy sign, we decide if the evolution---compatible with the boundary data---is classically allowed or forbidden. The dihedral angles are symplectically conjugate momenta to the length variables, making their analogy to the particle momenta less surprising. Nonetheless, it is remarkable that the tunneling trajectory is characterized by the analytic continuation of these momenta to complex values. The transition amplitude dominated by these tunneling trajectories is exponentially suppressed, and the suppression is provided by the classical action evaluated on the tunneling trajectory. 

We find one more suggestive analogy following \cite{Barrett1994semiclassical}. We can interpret the imaginary dihedral-angle momenta as a Wick rotation of the theory, $t \to i \tau$, commonly understood as a signature change. At the same time, we interpret the classically forbidden evolution of quantum geometries not as Euclidean tetrahedral dynamics but as gluing in a Lorentzian tetrahedron with a spacelike boundary, and this can be understood as a (brief and local) change of spacetime signature.

\medskip 
We have disentangled the question of tunneling of quantum geometries from the complication of spinfoam theories by considering the simplest model available, the Ponzano-Regge model. In doing so, we have provided the first concrete analysis of tunneling processes in spinfoam theory. We leave to future exploration the quantitative study of the classically forbidden trajectories in Regge calculus and their connection with gravitational instantons \cite{DonaHalFuture}. A similar scenario should also be realized in more physically appealing spinfoam theories for four-dimensional Lorentzian gravity. Preliminary studies in this direction show very promising applications to cosmology \cite{Bianchi:2010zs,Vidotto:2011qa,Asante:2021phx,Dittrich:2023rcr,Han:2024ydv} and black holes  \cite{Christodoulou:2016vny, DAmbrosio:2020mut, Christodoulou:2018ryl, Soltani:2021zmv, Christodoulou:2023psv}.

This is a first step in the direction of more fully understanding and characterizing purely quantum gravitational processes, and there are still many open questions. Tunneling processes of quantum geometries will soon play a major role in the search for physical signatures of covariant Loop Quantum Gravity. 
A natural next step is to understand how to compute tunneling probabilities and physically measurable quantities (e.g., halve-lives) in terms of the spinfoam amplitude. 

\section{Acknowledgments}
This work was made possible through the support of the ID\# 62312 grant from the John Templeton Foundation, as part of the project \href{https://www.templeton.org/grant/the-quantum-information-structure-ofspacetime-qiss-second-phase}{``The Quantum Information Structure of Spacetime'' (QISS)}. The opinions expressed in this work are those of the authors and do not necessarily reflect the views of the John Templeton Foundation. 
C.R. acknowledges support from the Perimeter Institute for Theoretical Physics through its distinguished research chair program. H.M.H. and F.V. acknowledge support from the Perimeter Institute for Theoretical Physics through its affiliation program. Research at Perimeter Institute is supported by the Government of Canada through Industry Canada and by the Province of Ontario through the Ministry of Economic Development and Innovation.
Research in F.V.'s research group at Western University is supported by the Canada Research Chairs Program and by the Natural Science and Engineering Council of Canada (NSERC) through the Discovery Grant ``Loop Quantum Gravity: from Computation to Phenomenology."  
We acknowledge the Anishinaabek, Haudenosaunee, L\=unaap\'eewak, Attawandaron, and neutral peoples, on whose traditional lands Western University and the Perimeter Institute are located.


\onecolumngrid
\appendix

\section{Formulas for the geometry}
\label{app:geo}
This appendix collects a few useful geometrical formulas for Euclidean tetrahedra. The dynamical variables in Regge gravity are the lengths of the edges of the Euclidean tetrahedra. 

We can derive all the geometrical quantities of the tetrahedron as a function of the lengths starting from the Cayley-Menger matrix
\begin{equation}
    C = \left(\begin{array}{ccccc}
        0 & 1           & 1           & 1           &        1    \\
        1 & 0           & \ell_{12}^2 & \ell_{13}^2 & \ell_{23}^2 \\
        1 & \ell_{12}^2 & 0           & \ell_{14}^2 & \ell_{24}^2 \\
        1 & \ell_{13}^2 & \ell_{14}^2 & 0           & \ell_{34}^2 \\
        1 & \ell_{23}^2 & \ell_{24}^2 & \ell_{34}^2 & 0           \\
    \end{array}\right) \ .
\end{equation}
The squared volume of the tetrahedron is given by $V^2(\ell_{ab}) = \frac{1}{144}\det C$. The squared area of any triangle is given by  $S_a^2(\ell_{ab}) = -\frac{1}{16}\det C_{a}$, where $C_{a}$ is the minor of $C$ where we in which the row and column with the lengths not involving $a$ have been eliminated. (For example, to get the area of the triangle $1$ we eliminate the last row and column).
The formula for $S_a^2(\ell_{ab})$ is just Heron's formula for the area of an Euclidean triangle, and the formula for $V^2(\ell_{ab})$ was first derived by Piero della Francesca in the 15th century. The advantage of using the Cayley-Menger matrix in 3D is marginal, but this matrix can be readily generalized to arbitrary dimensions, a valuable feature.

\medskip

The external dihedral angle $\psi_{ab}$ dual to the length $\ell_{ab}$ involves the volume of the tetrahedron and the area of the two triangles that have $\ell_{ab}$ as edge
\begin{equation}
\label{eq:angles}
    \sin  \psi_{ab}(\ell_{ab}) = \frac{3}{2}\frac{\sqrt{V^2(\ell_{ab})} \ell_{ab}}{S_a(\ell_{ab}) S_b(\ell_{ab})} \ .
\end{equation}
(Here, we have left $\sqrt{V^2}$ unsimplified to emphasize that $V^2<0$ arises in the continuation of this formula to the Lorentzian case.) This formula expresses the sine of the dihedral angle as a function of just the lengths.

For a classically forbidden geometry, $V(\ell_{ab})$ is purely imaginary. Since $\ell_{ab}>0$ and $S_a(\ell_{ab})>0$ necessarily $ \sin \psi_{ab}(\ell_{ab})$ is also purely imaginary. We can express the analytic continuation of the Euclidean dihedral angle in terms of the (real) Lorentzian boost angles $\psi^L_{ab}$ as
\begin{equation}
   \psi_{ab}(\ell_{ab}) =   \chi_{ab}(\ell_{ab}) + i \psi^L_{ab}(\ell_{ab})   \ ,
\end{equation}
where the factor $\chi_{ab}(\ell_{ab}) =0$ or $\pi$ and the sign of $\psi^L_{ab}(\ell_{ab})$ must be fixed depending on whether the dihedral angle is co-chronal or anti-chronal \cite{Sorkin2019, Dona:2020yao, Asante2021}.

\newpage
\twocolumngrid
\bibliography{biblio.bib}
\bibliographystyle{bib-style}

\end{document}